\documentclass[a4paper]{spie}  

\usepackage{amsmath,amsfonts,amssymb}
\usepackage{graphicx}
\usepackage[left=1.925cm, right=1.925cm, top=2.54cm, bottom=4.94cm]{geometry}
\usepackage[colorlinks=true, allcolors=blue]{hyperref}
\usepackage[position=top]{subfig}
\usepackage{floatrow}

\title{Wakefield Excited by Ultrashort Laser Pulses in Near-Critical Density Plasmas}

\author[a, b]{Petr Valenta}
\author[a, b]{Ondrej Klimo}
\author[b]{Gabriele M. Grittani}
\author[c]{Timur Zh. Esirkepov}
\author[b]{Georg Korn}
\author[b, c]{Sergei V. Bulanov}
\affil[a]{Czech Technical University in Prague, Faculty of Nuclear Sciences and Physical Engineering, Prague, Czech Republic}
\affil[b]{ELI Beamlines, Institute of Physics, Czech Academy of Sciences, Prague, Czech Republic}
\affil[c]{Kansai Photon Science Institute, National Institutes for Quantum and Radiological Science and Technology, Kizugawa, Kyoto, Japan}

\authorinfo{Further author information: (Send correspondence to P.V.)\\P.V.: E-mail: petr.valenta@eli-beams.eu}

\begin{document} 
\maketitle

\begin{abstract}
Laser wakefield acceleration (LWFA) using high repetition rate mJ-class laser systems brings unique opportunities for a broad range of applications. In order to meet the conditions required for the electron acceleration with lasers operating at lower energies, one has to use high density plasmas and ultrashort pulses. In the case of a few-cycle pulse, the dispersion and the carrier envelope phase effects can no longer be neglected. In this work, the properties of the wake waves generated by ultrashort pulse lasers in near-critical density plasmas are investigated. The results obtained may lead to enhancement of the quality of LWFA electron beams using kHz laser systems.
\end{abstract}

\keywords{Laser wakefield accelerator, high repetition rate plasma accelerator, single-cycle laser pulse, dispersion, carrier envelope phase}

\section{INTRODUCTION}
\label{sec:intro}  

Laser wakefield acceleration (LWFA), whose idea stems from the paper by Tajima and Dawson \cite{Tajima1979}, is currently a well-established technique for producing high-energy electron bunches in a plasma medium. Within this concept, a relativistically intense laser pulse propagating through underdense plasma induces a strong wakefield which, in turn, may accelerate electrons \cite{Esarey1996, Esarey2009, Hooker2013, Bulanov2016}. In fact, ionized plasmas can sustain electric fields several orders of magnitude larger than conventional radio-frequency accelerators \cite{Dawson1959}, allowing one to substantially reduce the acceleration length. This pawes a way towards a new generation of compact electron accelerators.

Over the last few decades, the quality of accelerated electron beams has rapidly evolved mainly due to the advances in technology and better understanding of the underlying physics. First high-quality, quasi-monoenergetic electron beams were obtained in 2004 \cite{Mangles2004, Geddes2004, Faure2004} operating the LWFA experiments in the blowout (or bubble) regime \cite{Pukhov2002, Lu2006, Lu2007} in which the wakefield takes the form of an ion cavity surrounded by thin electron walls. In order to reach this regime, the laser pulse must have transverse and longitudinal size comparable with the plasma wavelength and its intensity must be sufficiently high to expel all the electrons from the vicinity of the propagation axis. Today, LWFA has demonstrated (although not simultaneously) the capability to produce electron bunches in the multi-GeV energy range with a relative energy spread of a few percent \cite{Kim2013, Leemans2014, Gonsalves2019}, a few femtosecond duration \cite{Tilborg2006, Ohkubo2007, Debus2010, Lundh2011} and hundreds of pC of charge \cite{Li2017}. These achievements make LWFA increasingly attractive for a wide range of multi-disciplinary experiments and applications (e.g. radiography \cite{Glinec2005}, radiotherapy \cite{Malka2010, DesRosiers2000}, radiolysis \cite{Malka2010, Gauduel2010}).

Most of the current LWFA experiments, however, are performed using Joule-class laser systems which, owing to constraints in laser technology, operate at low repetition rates. Consequently, generated electron sources lack in several important characteristics, such as shot-to-shot stability and signal-to-noise ratio, which limit their applicability \cite{Hooker2013}. Increasing the repetition rate opens up the possibility to significantly improve aforementioned parameters of electron sources and, in addition, enhance the average electron current by several orders of magnitude \cite{He2013a, He2013, He2016, Beaurepaire2015, Salehi2017, Guenot2017, Gustas2018, Faure2018}.

On the other hand, high repetition rate lasers are currently capable to deliver pulses only at multi-mJ energy level. According to the scaling laws of the blowout regime \cite{Gordienko2005, Lu2007, Bulanov2016}, downscaling the LWFA to the mJ level calls for the use of extremely short laser pulses and high density plasmas. Typically, for a 1 TW laser system delivering pulses at wavelength $ \lambda_0 = 0.8 \ \mathrm{\mu m} $, one requires pulse of duration $ \tau < 5 \ \mathrm{fs} $ focused down to a spot of size $ w_0 \approx 2 \ \mathrm{\mu m} $ in a plasma of electron density $ n_e \approx 10^{20} \ \mathrm{cm^{-3}} $ \cite{Faure2018}. Although such requirements are achievable at the state of the art \cite{Faure2018}, the interaction of few-cycle laser pulses with high-density plasmas at relativistic intensities is not yet fully explored. In fact, parameters of generated electron sources are very sensitive to nonlinear effects which take place during such interaction, therefore its proper understanding may lead to enhancement of the quality of LWFA electron beams at kHz repetition rate.

Previously, the propagation of few-cycle laser pulses through underdense plasmas \cite{Schmid2009} as well as the effects of carrier envelope phase \cite{Lifschitz2012} and the influence of chirp on the laser propagation \cite{Beaurepaire2014} were studied. Specificities of relativistic laser-plasma interaction using near-single-cycle laser pulses are described in detail in Ref.~\citenum{Faure2018}.

The present paper contains the results of studying the propagation of single-cycle laser pulse through near-critical density plasmas using computer simulations. We show that the dispersion and carrier envelope phase effects cannot be neglected and derive analytical formulas describing the properties of the wakefield excited by few-cycle laser pulses.

\section{DISPERSION AND CARRIER ENVELOPE PHASE EFFECTS}
\label{sec:dispersion}  

While a small amplitude electromagnetic wave propagates in the collisionless plasma, its frequency $ \omega $ dependence on wave number $ k $ is given by the dispersion equation~\cite{Ginzburg1964}
\begin{equation*}\label{disp-eq}
\omega^2 = k^2 c^2+\omega_{pe}^2,
\end{equation*}
where $ c $ is the speed of light in vacuum and $ \omega_{pe} = \sqrt{4\pi n_e e^2/m_e} $ is the Langmuir frequency. Here $ e $ and $ m_e $ are the electron electric charge and mass, respectively, and $ n_e $ is the electron density. The phase and group velocity of the laser pulse are equal to 
\begin{equation*}\label{v-ph}
v_{ph} = \frac{\omega}{k} = c \frac{\omega}{\sqrt{\omega^2 - \omega_{pe}^2}}
\end{equation*}
and
\begin{equation*}\label{v-g}
v_g = \frac{\partial \omega}{\partial k} = c \frac{\sqrt{\omega^2 - \omega_{pe}^2}}{\omega},
\end{equation*}
respectively. As we may see, $v_{ph} v_g=c^2$.

The wave packet with the wave numbers within a relatively narrow wavenumber band, $ k_0 - \Delta k < k < k_0 + \Delta k $, and $ \Delta k \ll k_0 $ during the time interval less than the typical dispersion time
\begin{equation}\label{t-disp}
t_{disp} = \pi \frac{\sqrt{\omega^2 - \omega_{pe}^2}}{\omega_{pe}^2} = \frac{\pi}{\omega} \left( \frac{\beta_w}{1 - \beta_w^2} \right) = \frac{\pi}{\omega} \gamma_w \sqrt{\gamma_w^{2} - 1}
\end{equation}
can be approximated by the function
\begin{equation*}\label{pulse}
A \left(x, t \right) = a_0 \frac{\sin{ \left(\Delta k( x - \beta_w c t)\right)}}{ \Delta k (x-\beta_w c t)} \cos{\left(k_0 \left( x -\frac{c t}{\beta_w }\right)\right) }.
\end{equation*}
Here the normalized group velocity is equal to 
\begin{equation*}\label{bw}
\beta_w = \frac{v_g}{c}= \frac{k_0 c}{\sqrt{k_0^2 c^2 + \omega_{pe}^2}}
\end{equation*}
and $ \gamma_w = 1 / \sqrt{1-\beta_w^2} = \omega / \omega_{pe} $. From Eq. (\ref{t-disp}) it follows that the dispersion time equals $ t_{disp} = \pi k_0 c / \omega_{pe}^2 $. The electromagnetic pulse width is $ l = \pi/\Delta k $.

According to the LWFA theory (e.g. see Refs.~\citenum{Esarey2009, Bulanov2016}), the wake wave is excited in the plasma behind the driver laser pulse by the force equal to 
\begin{equation*}\label{driv-force}
F \left(x, t \right) = -\frac{1}{2\gamma} \partial_x A^2 \left(x, t \right) = f \left(\frac{X}{\tilde{l}}\right) \sin(2 \xi + \phi)
\end{equation*}
with $ f \left(X/\tilde{l}\right) = a_0^2 \left(\sin \left(X/\tilde{l}\right) / \left(X/\tilde{l}\right) \right)^2 / 2 \tilde{l} $, $ \tilde{l} = \pi k_0 / \Delta k $,  $ \gamma = \sqrt{1 + A^2 + p^2} $. The electromagnetic potential $ A $ is normalized on $ m_e c^2 / e $, $ p $ is the electron momentum and $ \phi $ is the phase. Here and below we use normalized variables
\begin{equation*}\label{varnorm}
X = k_0 \left(x - \beta_w c t\right) \quad \mathrm{and} \quad \xi = k_0 \left(x - \frac{c t}{\beta_w} \right) = X + \varkappa_t k_0 c t \equiv \frac{X}{\beta_w^2} + \varkappa_x k_0 x,
\end{equation*}
where
\begin{equation}\label{kappa}
\varkappa_t  = \beta_w - {1}/{\beta_w} \quad \mathrm{and} \quad \varkappa_x = 1 - 1/\beta_w^2.
\end{equation}
The dispersion time given by Eq.~(\ref{t-disp}) is equal to $ t_{disp} = \pi k_0 c/\varkappa_t $.

The equations describing the wake wave in the limit well below the wavebreaking ($ \gamma = 1 $) are
\begin{equation*}\label{equmot}
\partial_X p = -E + F, \qquad \partial_X E = p,
\end{equation*}
where $ E $ is the electric field in the wake wave. Their solution for $ \varkappa_x, \varkappa_t \ll 1 $ reads
\begin{equation*}\label{equ-sol}
p + i E = e^{i X} \int_{0}^{X} f \left( \frac{X'}{\tilde{l}} \right) \sin \left(\frac{2}{\beta_w^2} X' + 2 \varkappa_x k_0 x + \phi \right) e^{-i X'} \mathrm{d} X'.
\end{equation*}
For ultrashort driver pulse we have
\begin{equation*}\label{fin-sol-p}
p =\frac{a_0^2}{2} \sin \left( 2 \varkappa_x k_0x+ \phi \right) \cos \left( X\right),
\end{equation*}
\begin{equation*}\label{fin-sol-E}
E =\frac{a_0^2}{2} \sin \left( 2 \varkappa_x k_0x + \phi \right) \sin \left( X\right).
\end{equation*}

\floatsetup[figure]{style=plain, subcapbesideposition=top}
\begin{figure}[h!]
	\centering
	\sidesubfloat[]{\includegraphics[height=5cm]{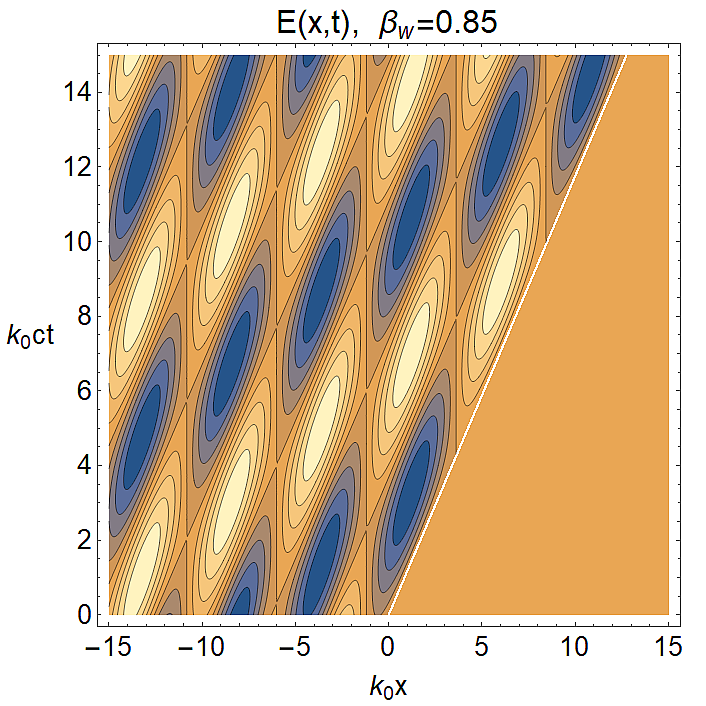}} \hspace{4mm}
	\sidesubfloat[]{\includegraphics[height=5cm]{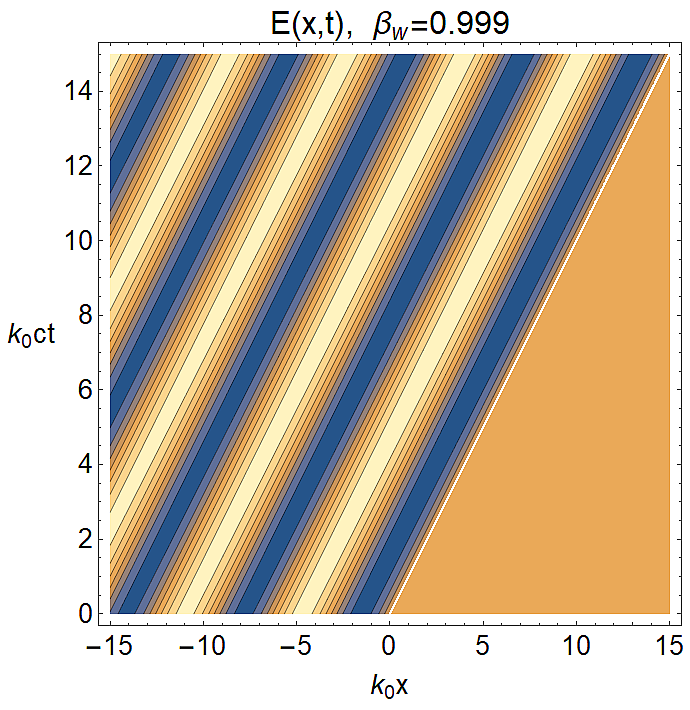}}
	\caption[theory]{\label{fig:theory} Contour-plots showing the wakefield, $ E(x,t) $, distribution in the $ (x,t) $ plane. The normalized group velocity of laser pulse is equal to (a) $ \beta_w = 0.85 $ and (b) $ \beta_w = 0.999 $.}
\end{figure}

Fig.~\ref{fig:theory} shows two contour-plots presenting the wakefield, $ E(x,t) $, distribution in the $ (x,t) $ plane for the normalized laser group velocity (it is the wake wave phase velocity) equal to $ \beta_w = 0.85 $ and $ \beta_w = 0.999 $, respectively. As we may see, in Fig.~\ref{fig:theory} (a) the dispersion effects result in the space and time modulations with the modulation period determined by the parameters $ \varkappa_x $ and $ \varkappa_t $ (see Eq.~(\ref{kappa})). In this case, $ \beta_w = 0.85 $ and the parameters are equal to $ \varkappa_x \approx 0.384 $ and $ \varkappa_t \approx 0.327 $, respectively. Fig.~\ref{fig:theory} (b) shows the wakefield pattern for $ \beta_w = 0.999 $ corresponding to the parameters $ \varkappa_x \approx 0.002 $ and $ \varkappa_t \approx 0.002 $. Here the modulations are not seen because their period is much longer.

\section{PARTICLE-IN-CELL SIMULATIONS}
\label{sec:simulations}  

The effects of dispersion and carrier envelope phase on the single-cycle laser pulse interaction with near-critical density plasmas are studied numerically by means of particle-in-cell (PIC) simulations in two-dimensional (2D) Cartesian geometry. The simulations are carried out using the fully relativistic electromagnetic PIC code EPOCH \cite{Arber2015}.

\subsection{Simulation Setup}

The simulated laser pulse, characterized by angular frequency $ \omega_0 = 2 \pi c / \lambda_0 $, where $ \lambda_0 $ is the vacuum center laser wavelength, is Gaussian in both spatial and temporal profiles. The pulse contains single optical cycle, thus its full width at half maximum (FWHM) duration is $ \tau = 1 \ T_0 $, where $ T_0 = 2 \pi / \omega_0 $ is the period of the laser pulse. The beam waist (half at $ 1/e $ of the vacuum laser electric field amplitude $ E_0 $) at the focal spot is $ w_0 = 4 \ \lambda_0 $, therefore the Rayleigh length is $ x_R = 16 \pi \lambda_0 $. The corresponding normalized amplitude at focus, defined as $ a_0 = e E_0 / (m_e \omega_0 c) $, is $ a_0 = 2 $. The laser beam is circularly polarized.

\begin{figure} [h!]
	\centering
	\includegraphics[height=4.5cm]{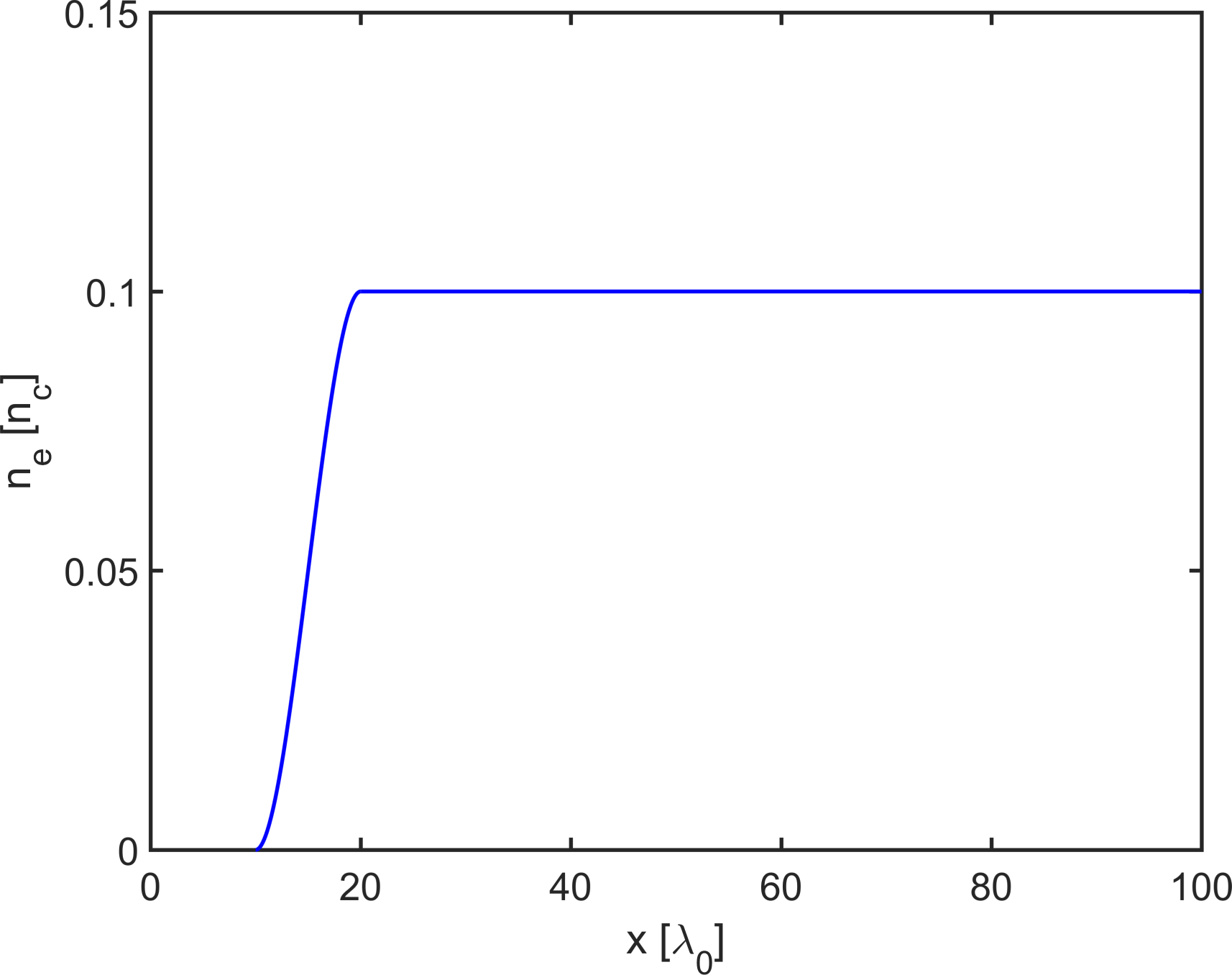}
	\caption[density] {\label{fig:density_profile} Longitudinal line-out of initial electron density profile.}
\end{figure}

The laser pulse propagates along the x-axis through a pre-ionized uniform hydrogen plasma with electron density $ n_e = 0.1 \ n_c $, where $ n_c = m_e \omega_0^2 / (4 \pi e^2) $ stands for the critical plasma density. A smooth ramp is added to the front side of the target in order to mitigate wavebreaking from sharp rising plasma edge \cite{Bulanov1990}. The ramp is defined by the following function, $ n_e \left( x \right) = 0.1 \ n_c \ (3 - 2 (x - x_1)/(x_2 - x_1)) ((x - x_1)/(x_2 - x_1))^2 $, where $ x_1 = 10 \ \lambda_0 $ and $ x_2 = 20 \ \lambda_0 $. Fig.~\ref{fig:density_profile} shows the longitudinal line-out of the electron density profile. The laser pulse is focused at the beginning of the density plateau, $ x_f = 20 \ \lambda_0 $, and the laser power should be well above the critical power for relativistic self-focusing \cite{Sun1987, Borisov1992} $ P_c = 17.4 \ n_c / n_e \ \mathrm{GW} $, therefore the self-focusing is expected to occur during the interaction. The plasma is cold, collisionless and is composed of electrons and protons represented by weighted macro-particles. The number of particles per one simulation cell is $ 100 $ for both electrons and protons.

The size of the simulation domain is $ 100 \ \lambda_0 $ in the laser propagation direction and $ 20 \ \lambda_0 $ in the transverse direction. The resolution of the 2D Cartesian grid is $ 100 $ cells per $ \lambda_0 $ in both longitudinal and transverse directions. The simulation time is set to $ 100 \ T_0 $. The electromagnetic fields are calculated using the standard second-order Yee solver \cite{Yee1966} and the interpolation between particles and grid nodes is achieved using a third order b-spline shape functions. The absorbing boundary conditions are applied on each of the simulation domain sides for both electromagnetic field and particles.

\subsection{Simulation Results}

\begin{figure}[h!]
	\centering
	\sidesubfloat[]{\includegraphics[height=3.5cm]{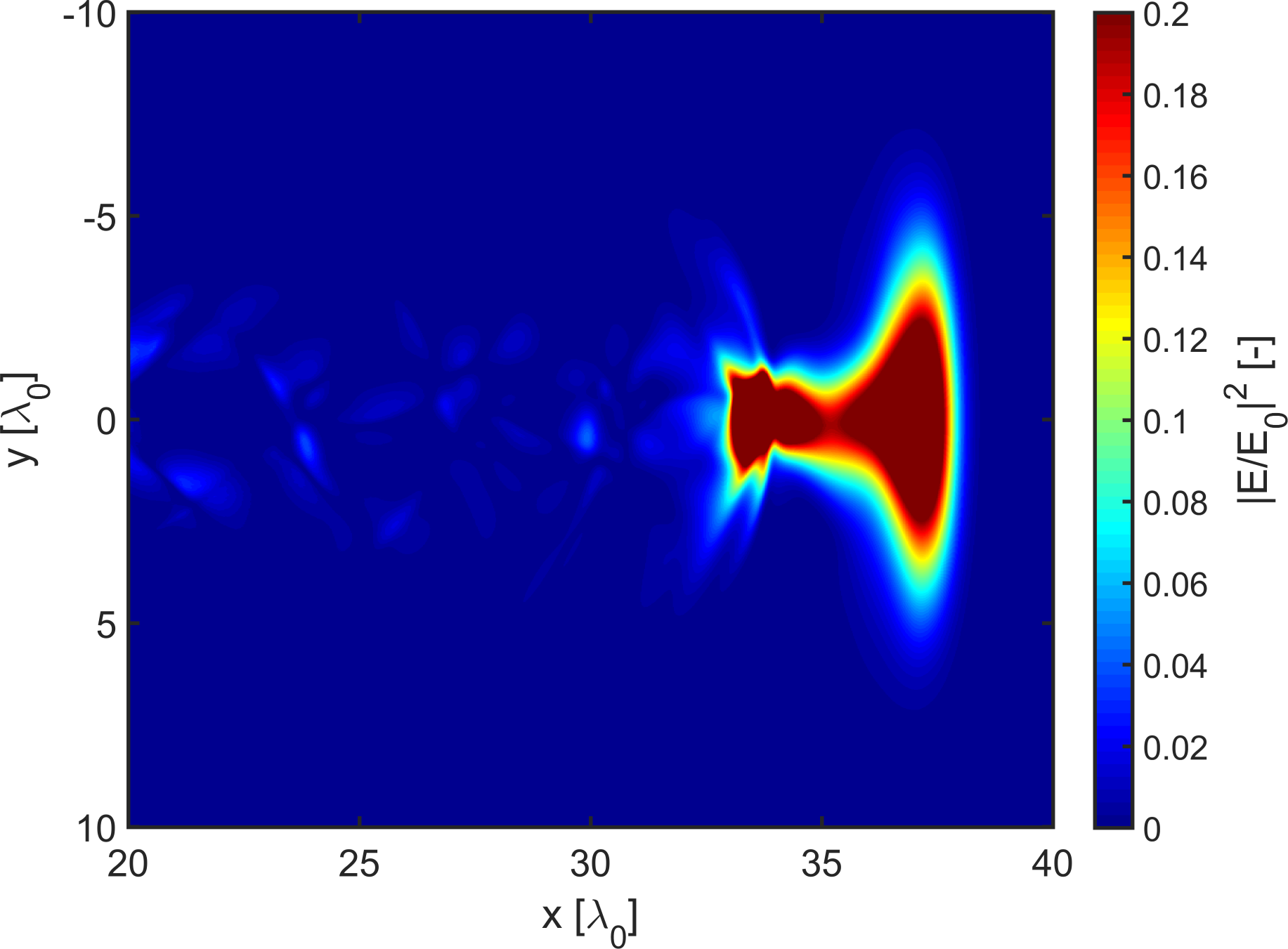}} \hspace{1mm}
	\sidesubfloat[]{\includegraphics[height=3.5cm]{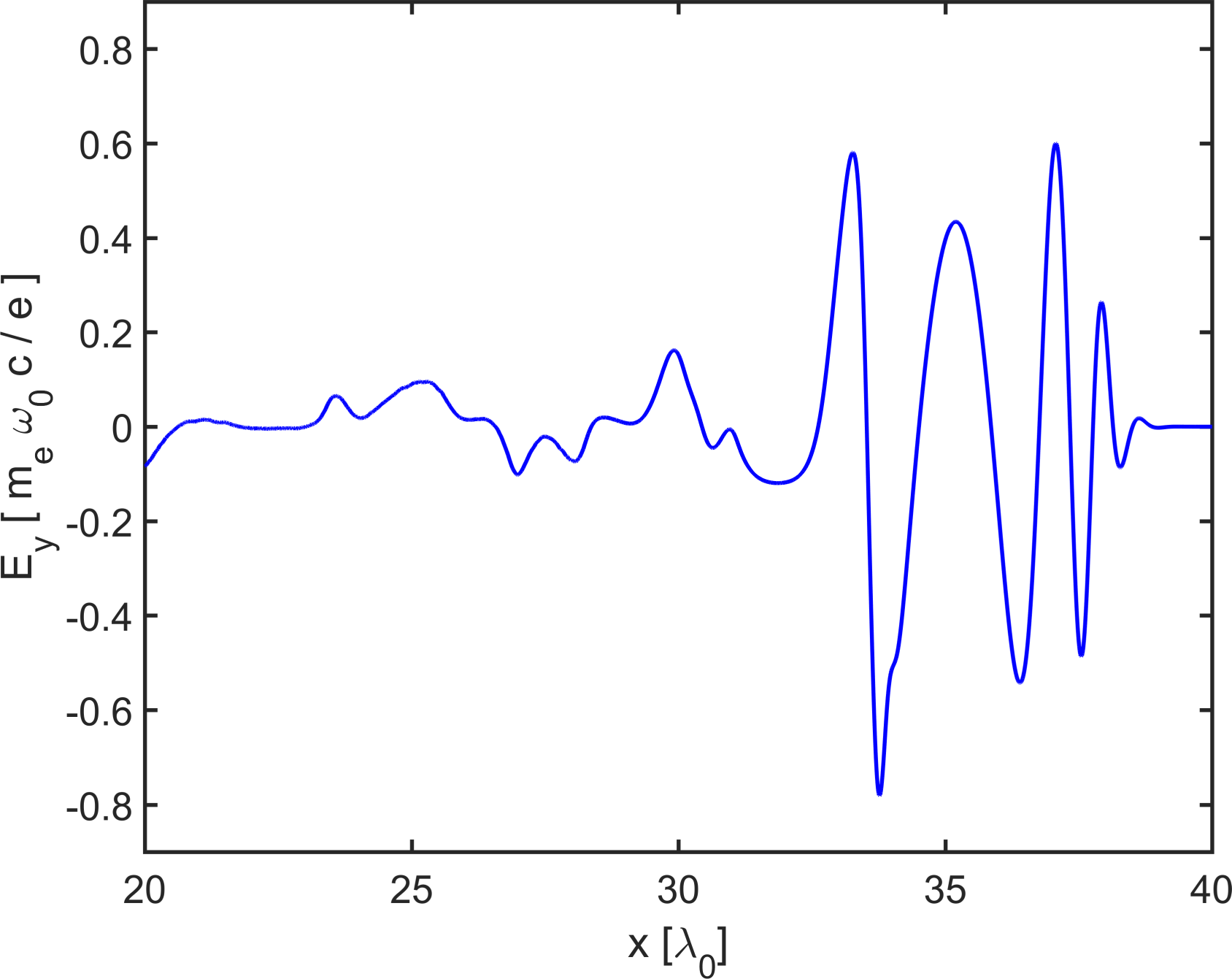}} \hspace{4mm}
	\sidesubfloat[]{\includegraphics[height=3.5cm]{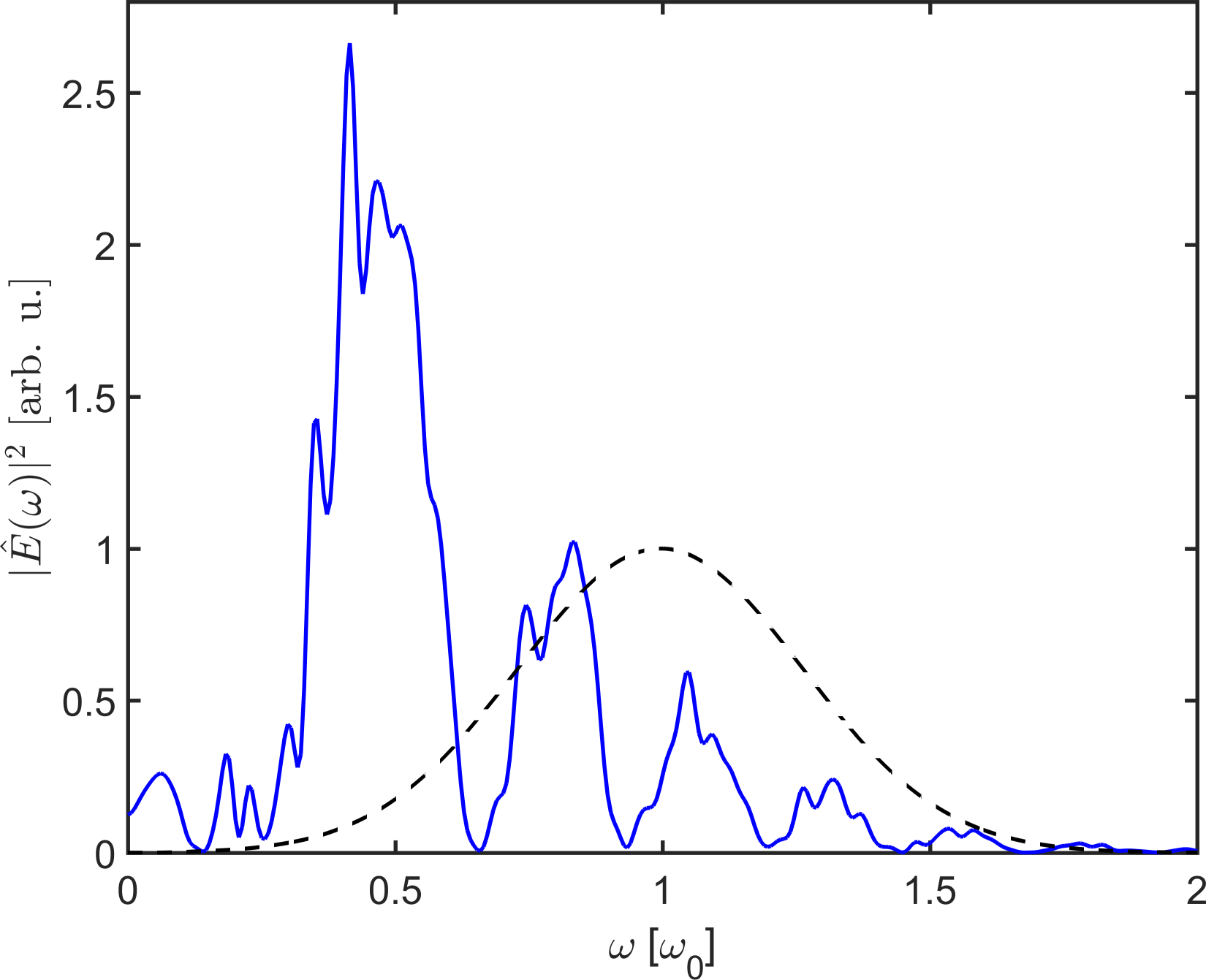}} \hspace*{5mm} \\
	\vspace{2mm} 
	\sidesubfloat[]{\includegraphics[height=3.5cm]{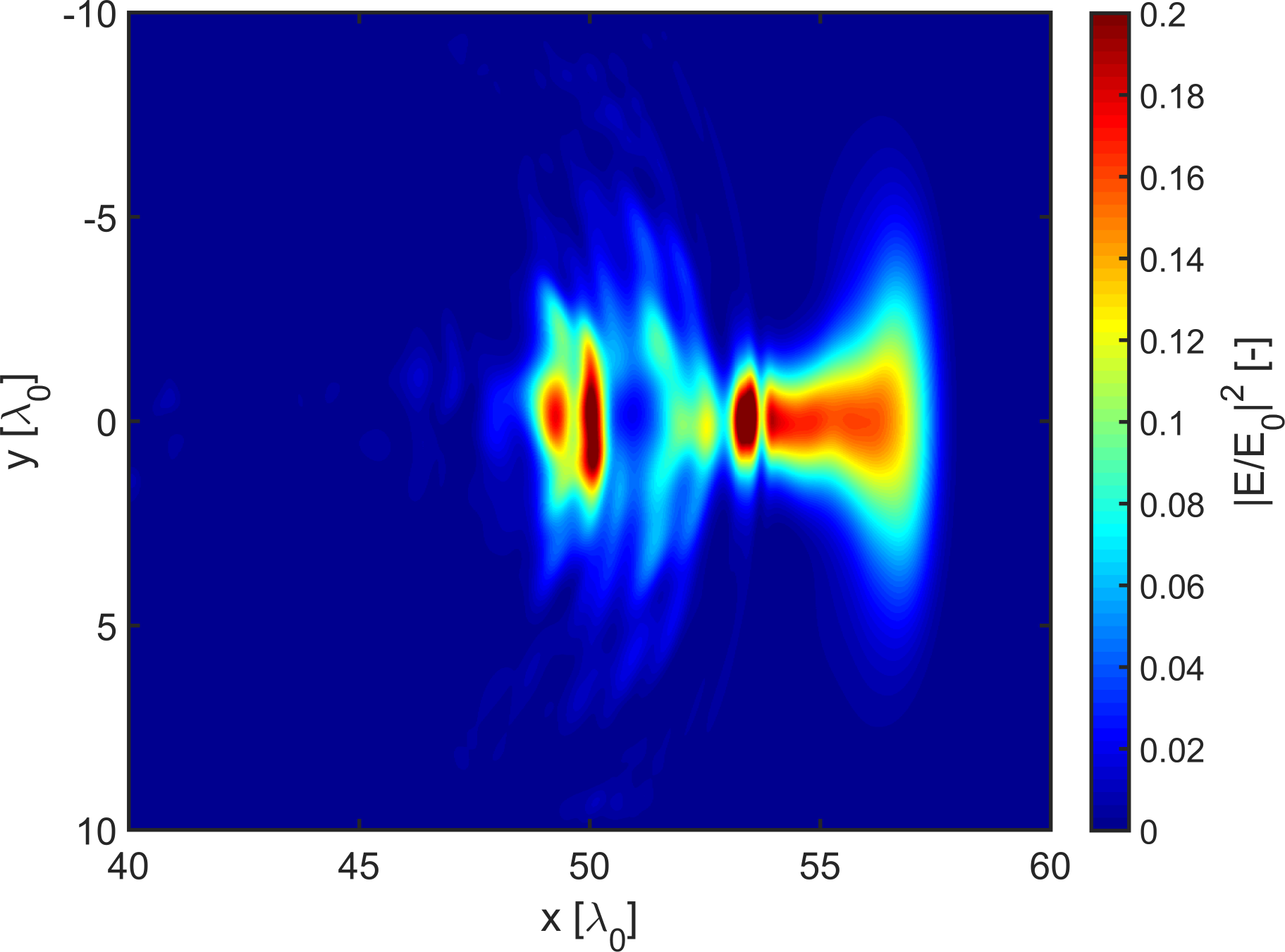}} \hspace{2mm}
	\sidesubfloat[]{\includegraphics[height=3.5cm]{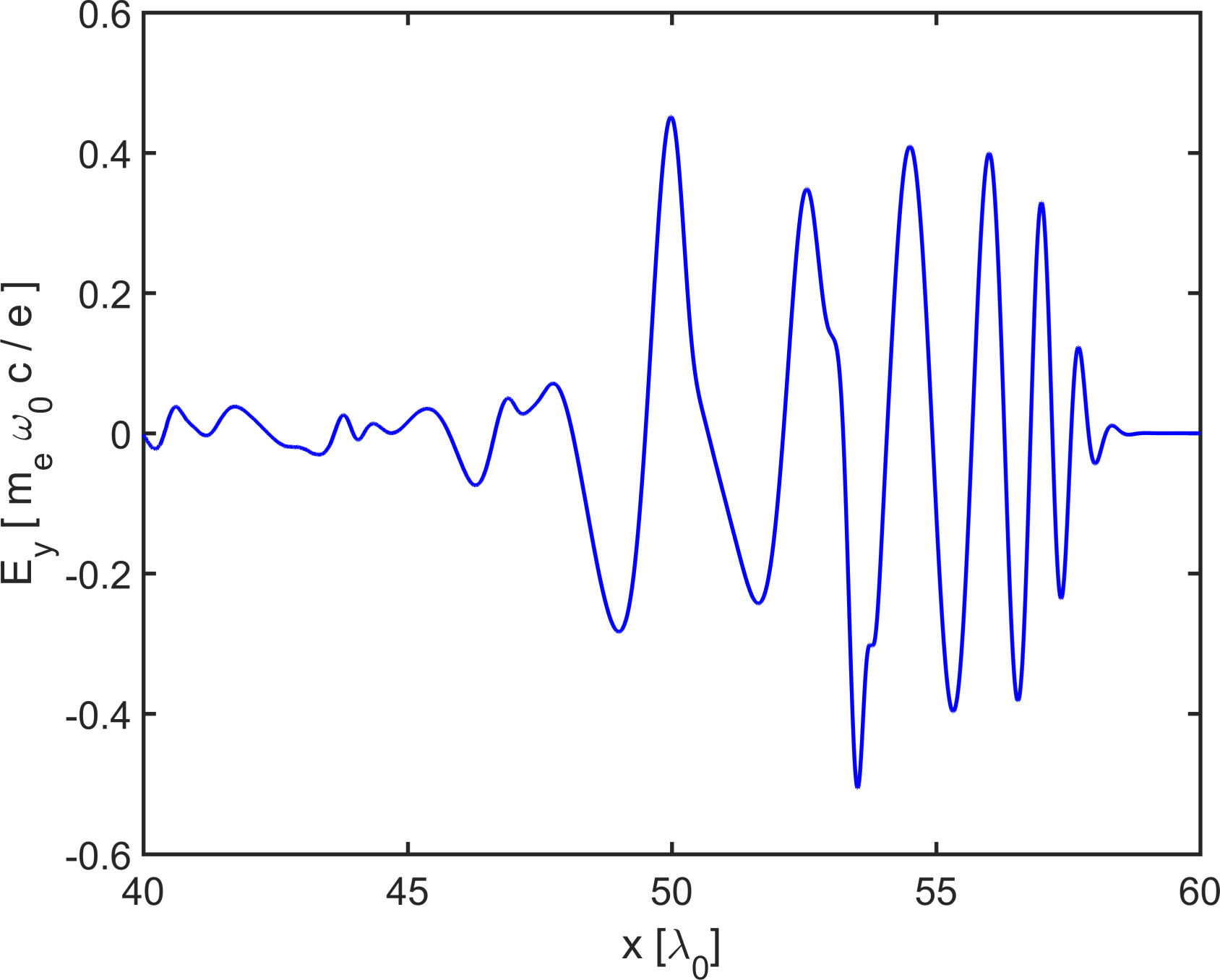}} \hspace{4mm}
	\sidesubfloat[]{\includegraphics[height=3.5cm]{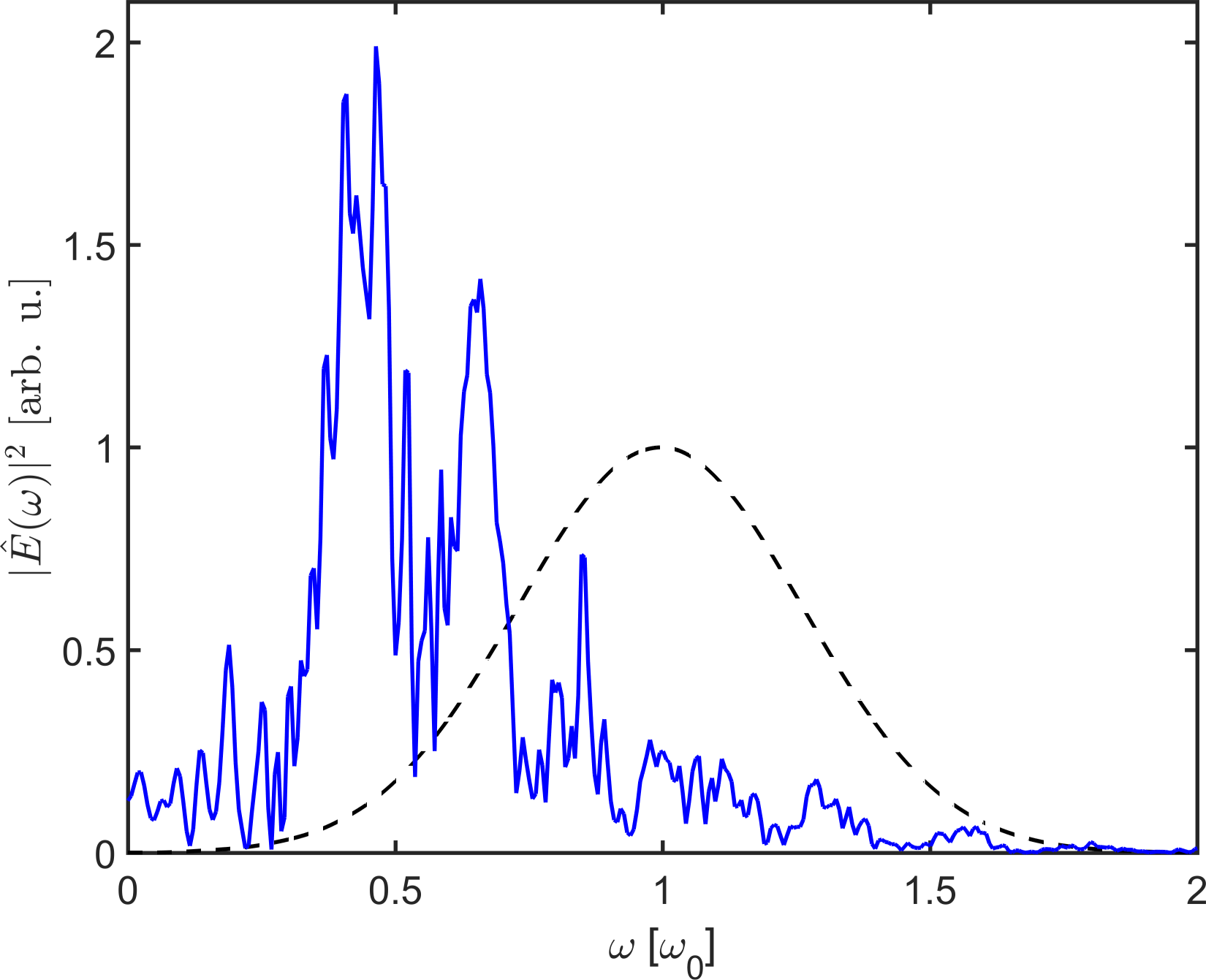}} \hspace*{4mm} \\
	\vspace{2mm}
	\sidesubfloat[]{\includegraphics[height=3.5cm]{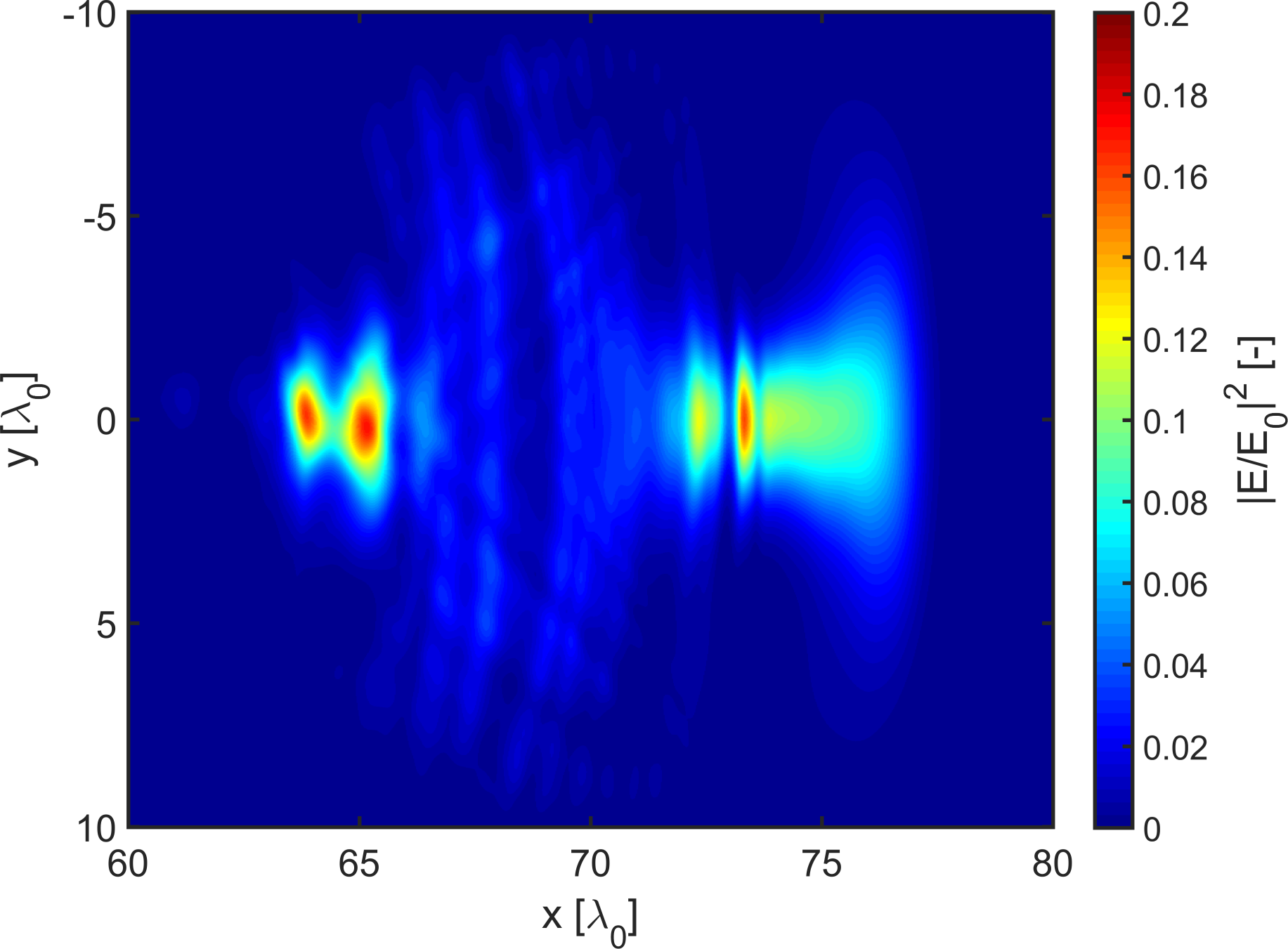}} \hspace{2mm}
	\sidesubfloat[]{\includegraphics[height=3.5cm]{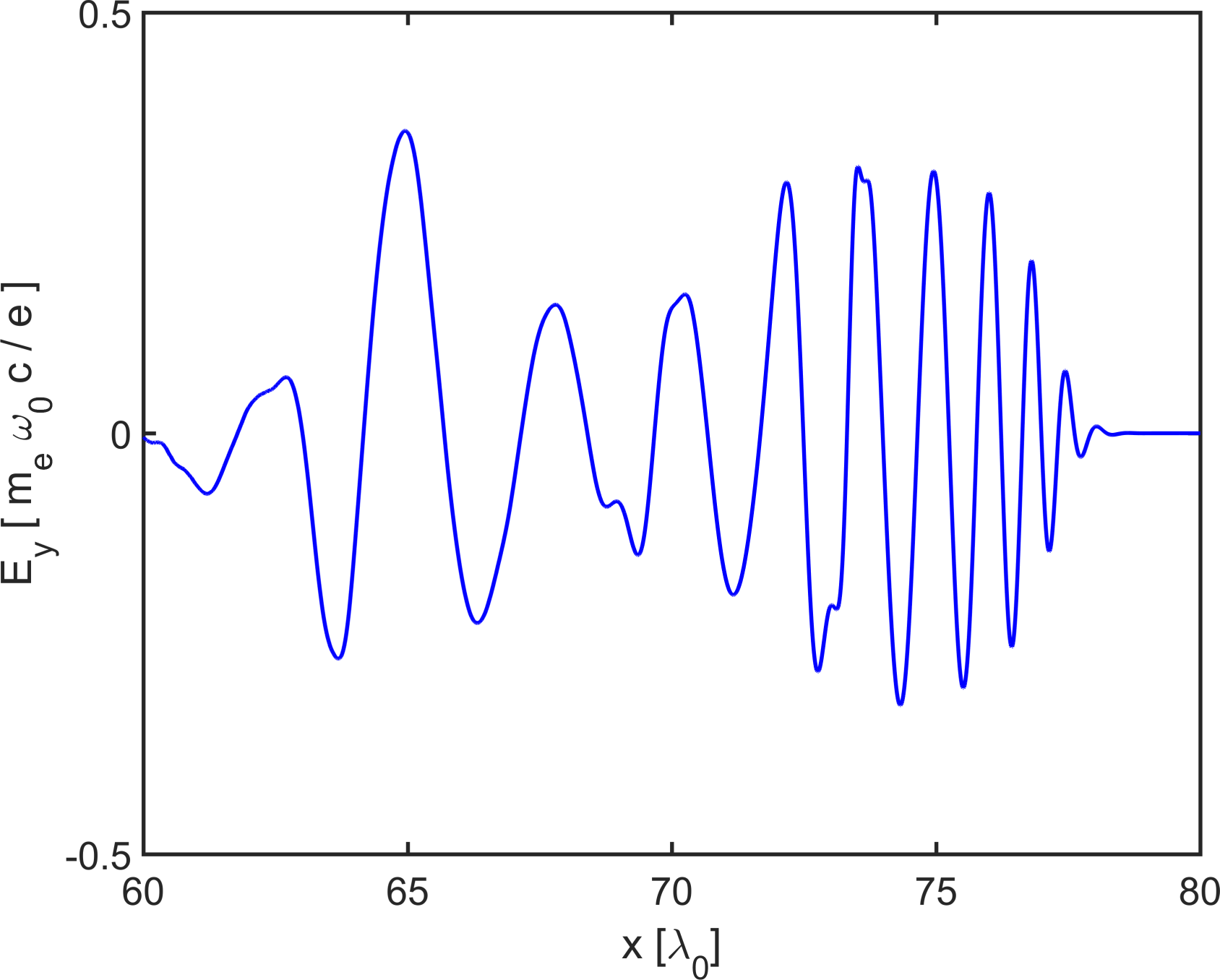}} \hspace{4mm}
	\sidesubfloat[]{\includegraphics[height=3.5cm]{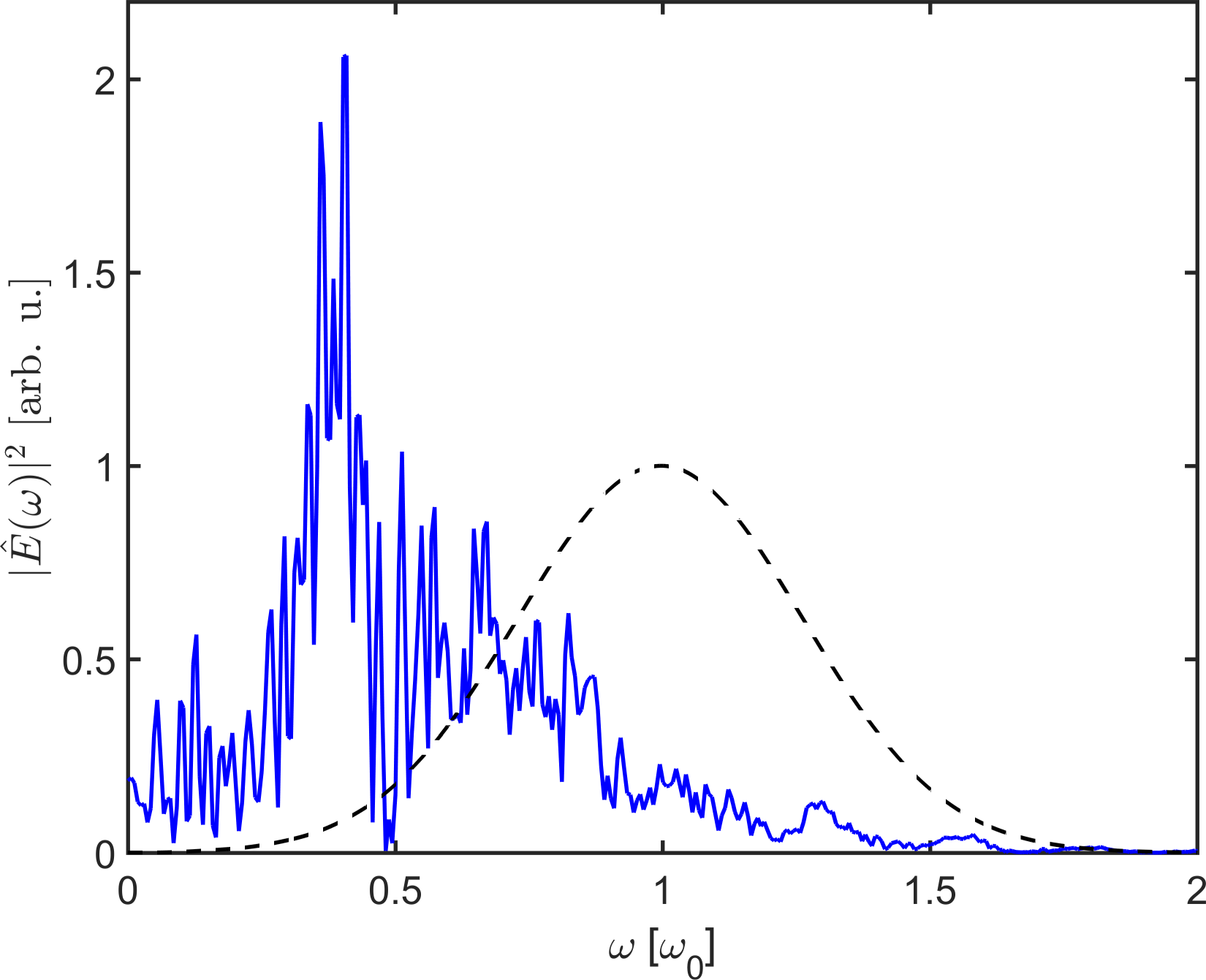}} \hspace*{4mm}
	\caption[2d]{\label{fig:laser_field} (a), (d), (g) Snapshots of the laser field $ |E/E_0|^2 $, (b), (e), (h) line-out along the axis $ y = 0 $ of the transverse electric field $ E_y $ and (c), (f), (i) the corresponding on-axis spectrum of the laser pulse at three successive time instants (a), (b), (c) $ t = 40 \ T_0 $, (d), (e), (f) $ t = 60 \ T_0 $ and (g), (h), (i) $ t = 80 \ T_0 $. The dashed lines in (c), (f), (i) represents the initial laser spectrum.}
\end{figure}

First, we analyze the evolution of the laser pulse in space and in time while it propagates through plasma. Fig.~\ref{fig:laser_field} shows snapshots of the laser field $ |E/E_0|^2 $, the transverse electric field $ E_y $ and the corresponding on-axis spectrum of the laser pulse at three different time instants. It can be clearly seen that the pulse evolution is very rapid and rather complex. As soon as the laser pulse enters high-density plasma region ($ x > 20 \ \lambda_0 $), it undergoes self-focusing (Fig.~\ref{fig:laser_field} (a)) and deposits its energy on the excitation of high-amplitude wake waves. Since the number of photons has to be roughly conserved during the interaction, the corresponding energy loss is manifested mainly through a frequency redshift which occurs at the front of the laser pulse \cite{Bulanov1992}.

\floatsetup[figure]{style=plain, subcapbesideposition=top}

\floatsetup[figure]{style=plain, subcapbesideposition=top}
\begin{figure}[h!]
	\centering
	\vspace{4mm}
	\sidesubfloat[]{\includegraphics[height=3.5cm]{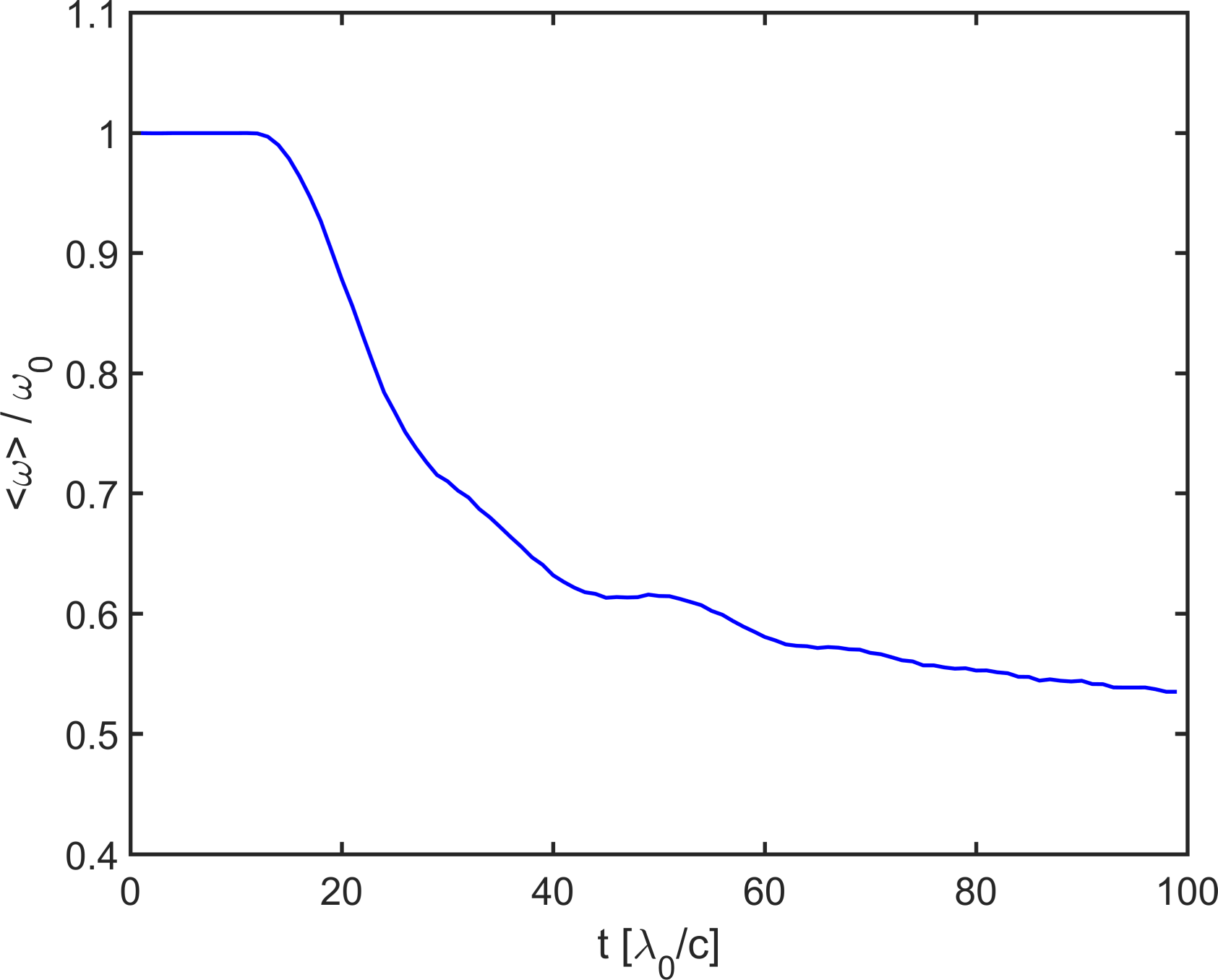}} \hspace{2mm}
	\sidesubfloat[]{\includegraphics[height=3.5cm]{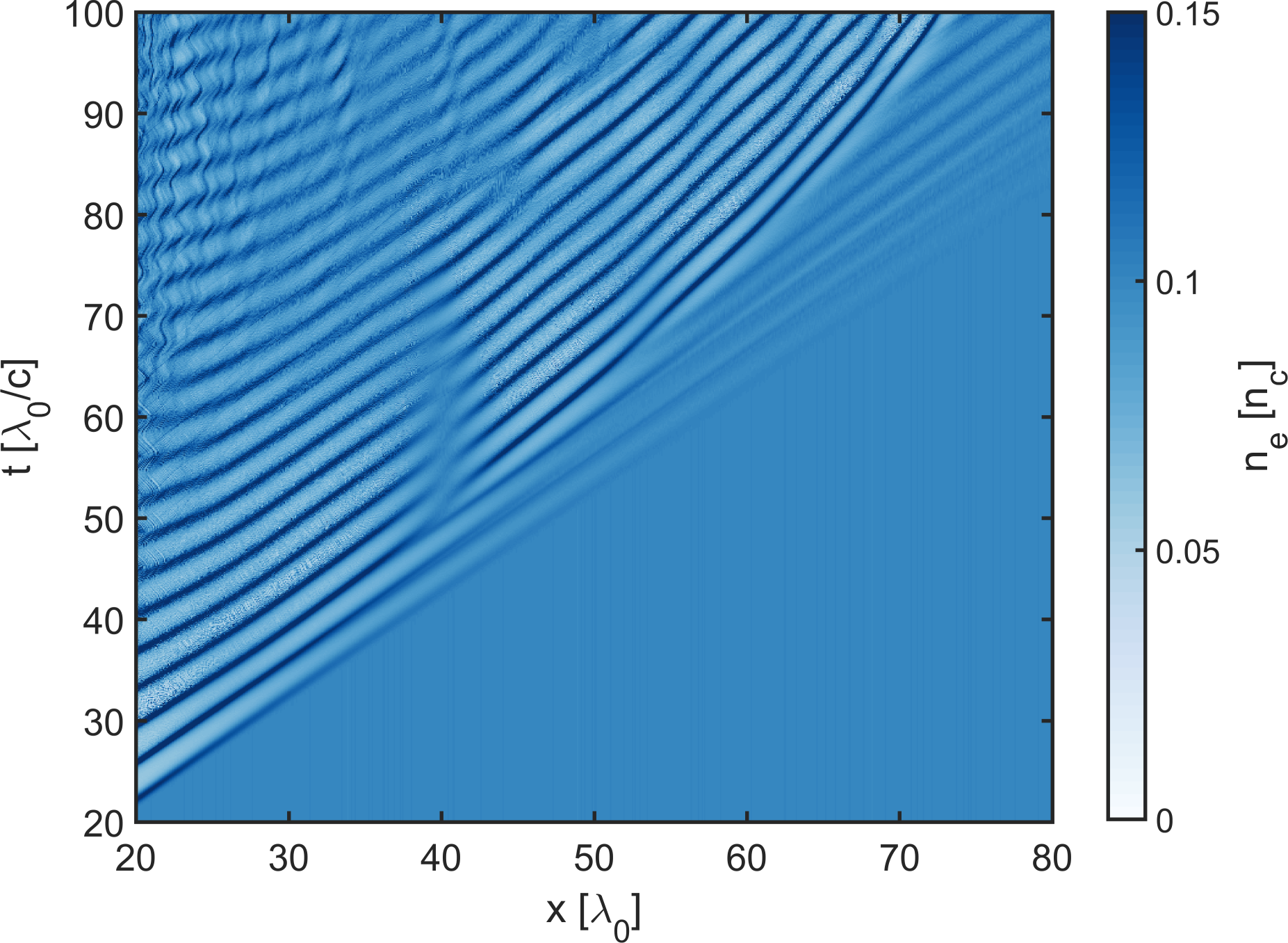}} \hspace{2mm}
	\sidesubfloat[]{\includegraphics[height=3.5cm]{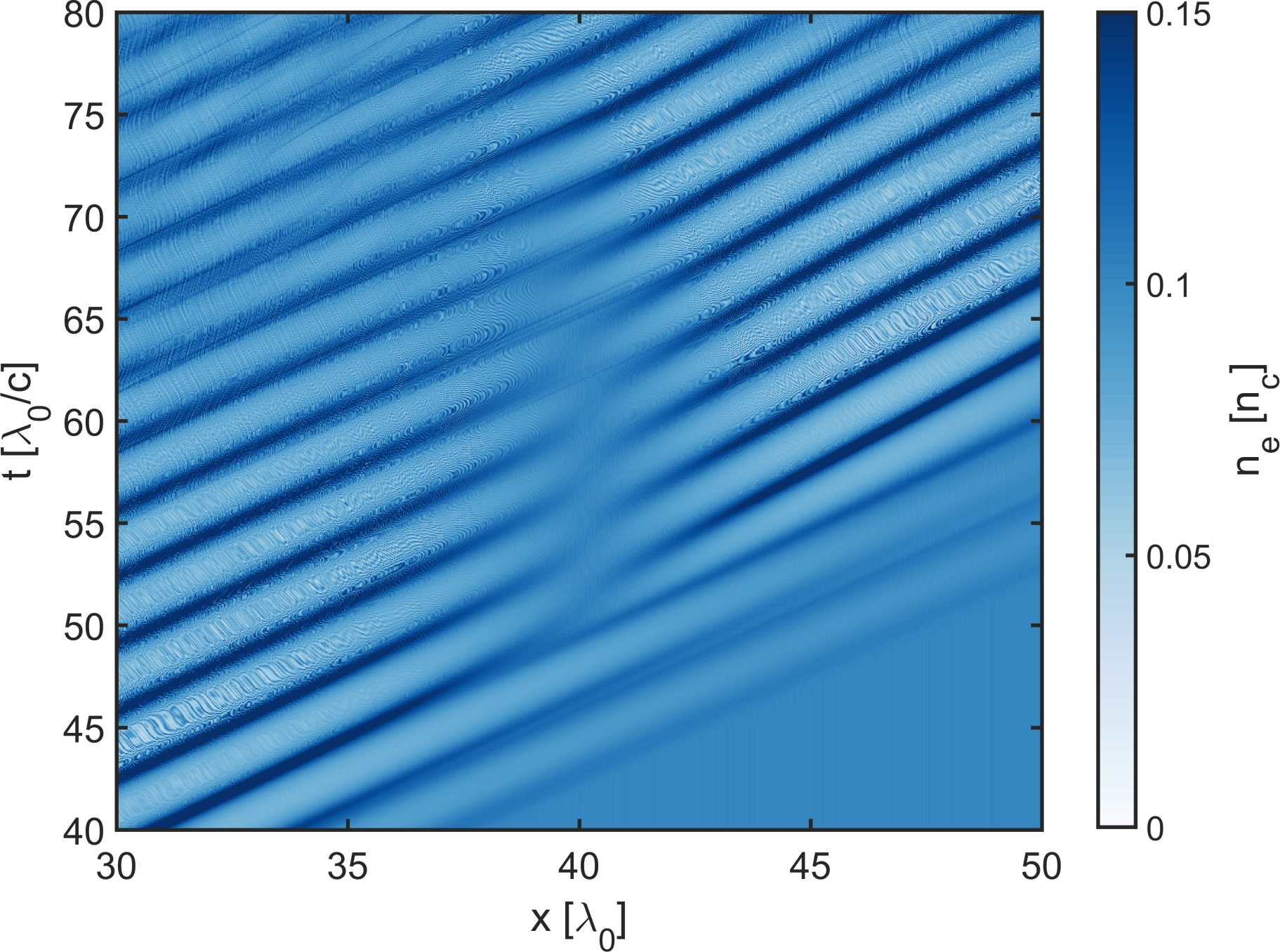}} 
	\caption[2d2]{\label{fig:density_evolution} (a) The average on-axis frequency $ \left\langle \omega \right\rangle  $ of the laser pulse in time, (b) the evolution of on-axis electron density $ n_e $ in the $ (x, t) $ plane and (c) a close-up view on the plasma density modulation of wake waves around $ x = 40 \ \lambda_0 $.}
\end{figure}

\floatsetup[figure]{style=plain, subcapbesideposition=top}
\begin{figure}[h!]
	\centering
	\sidesubfloat[]{\includegraphics[height=3.5cm]{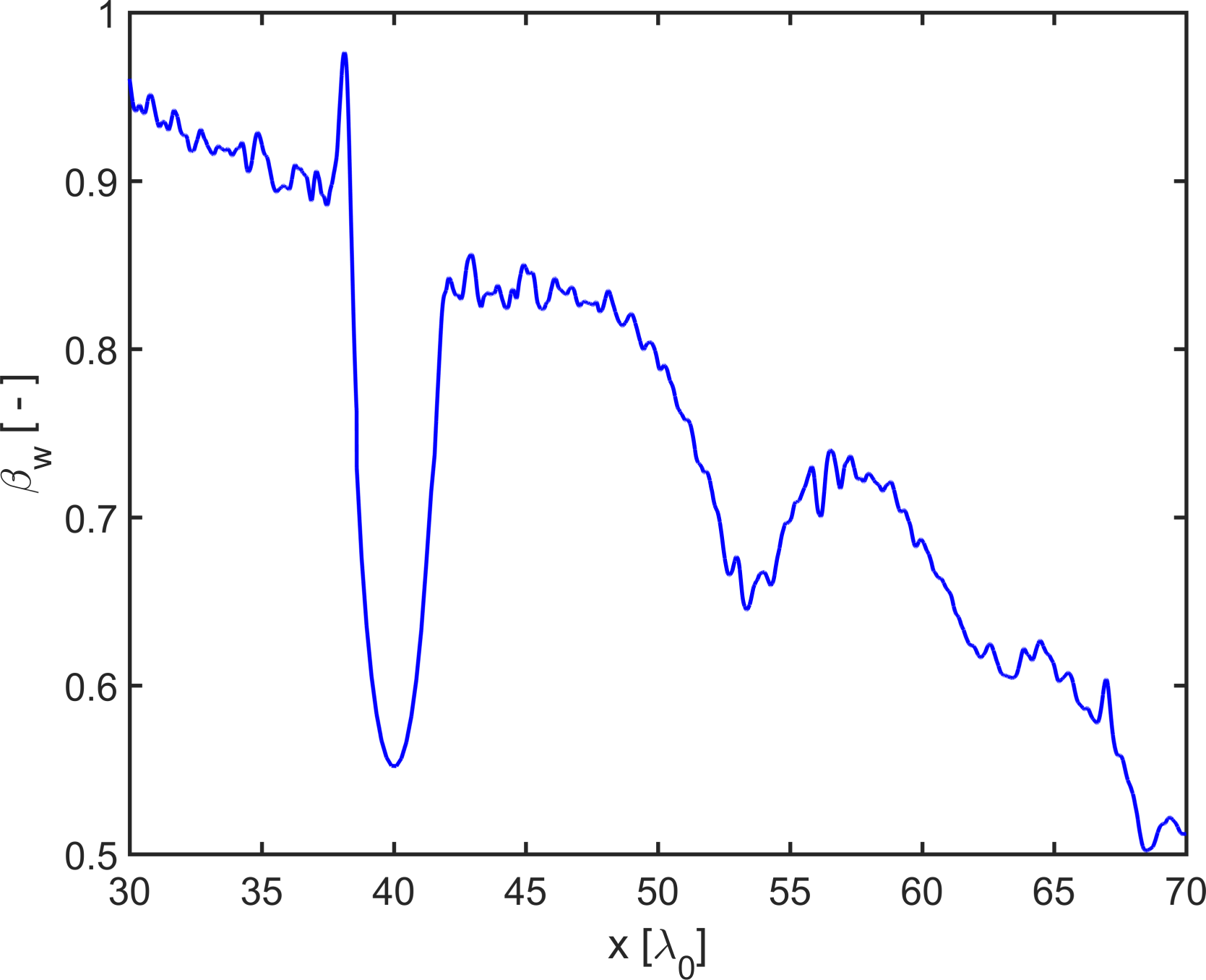}} \hspace{4mm}
	\sidesubfloat[]{\includegraphics[height=3.5cm]{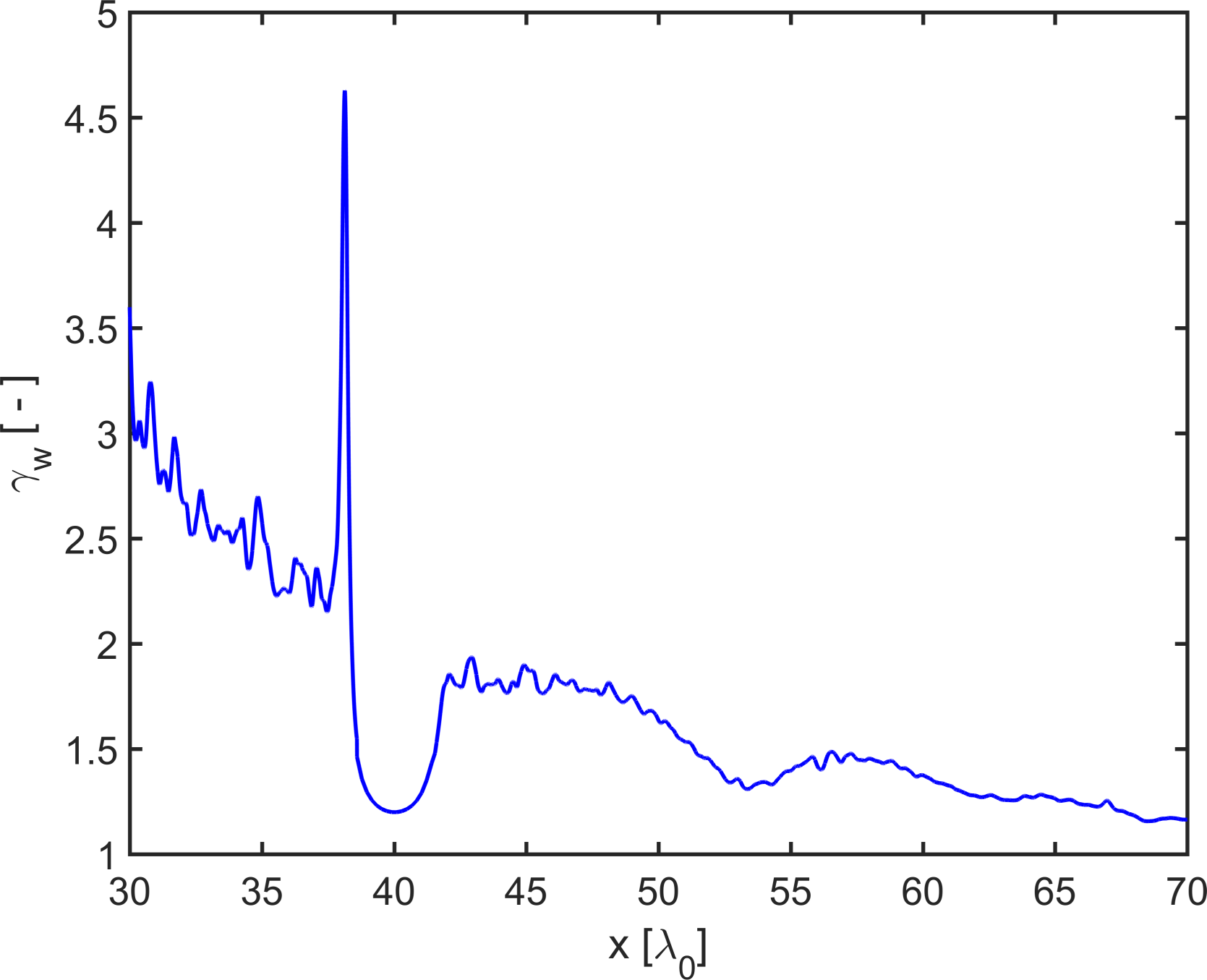}}
	\caption[1dlinear] {\label{fig:beta_and_gamma} Estimated time evolution of the (a) normalized on-axis phase velocity $ \beta_w $ of the density amplitude of one selected wake wave and (b) the corresponding $ \gamma_w $.}
\end{figure}

\floatsetup[figure]{style=plain, subcapbesideposition=top}
\begin{figure}[h!]
	\centering
	\vspace{3mm}
	\sidesubfloat[]{\includegraphics[height=3.5cm]{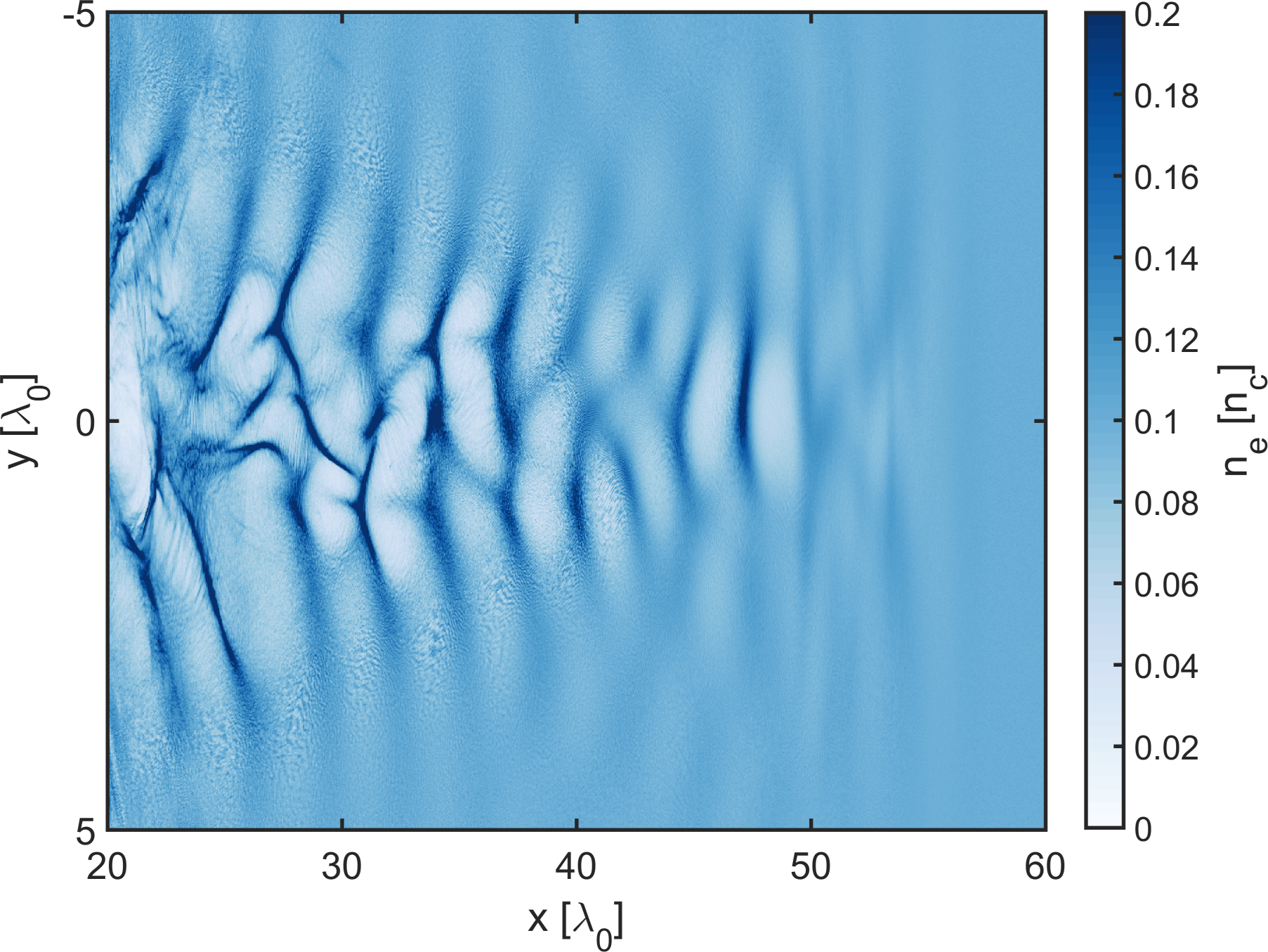}} \hspace{2mm}
	\sidesubfloat[]{\includegraphics[height=3.5cm]{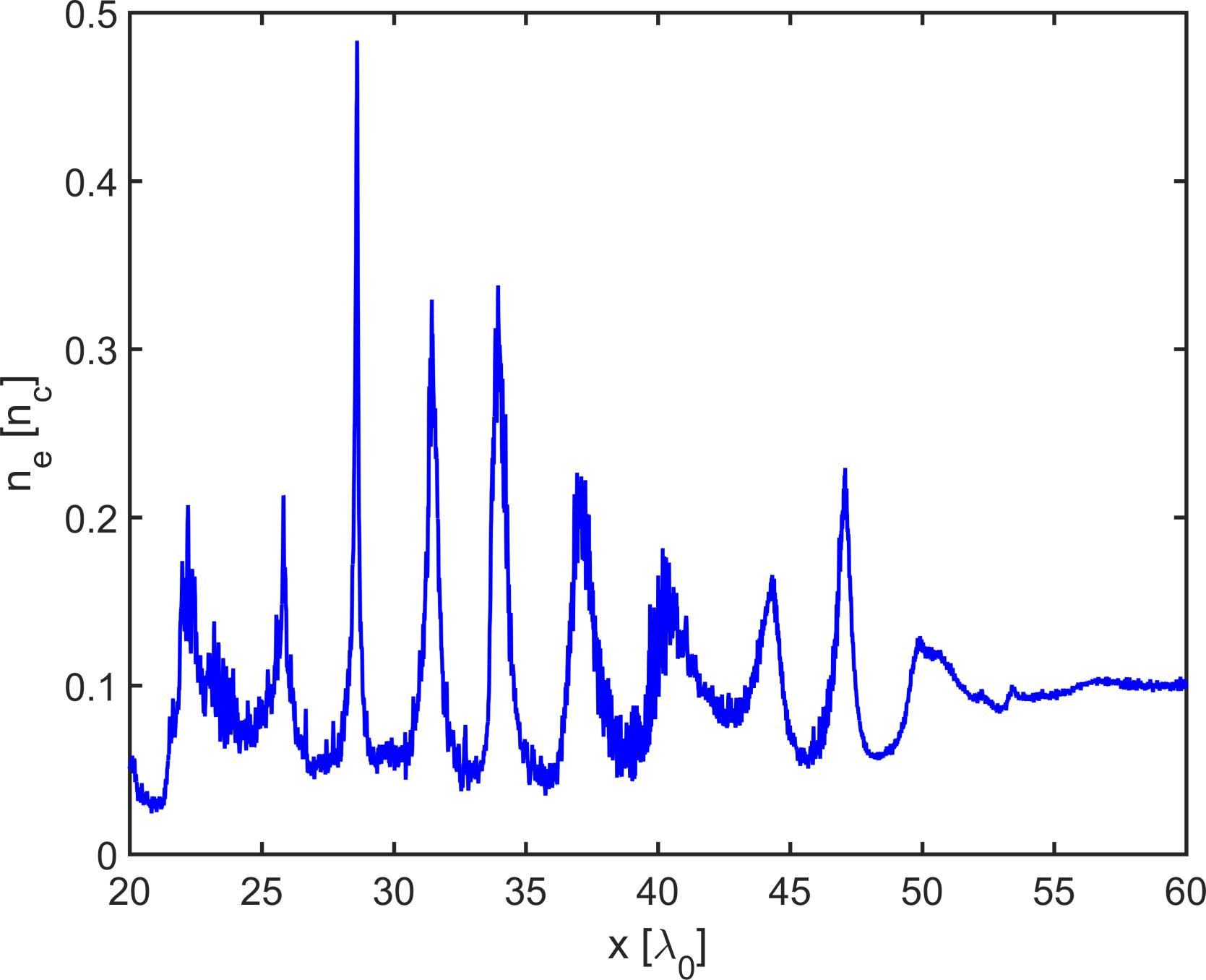}} \hspace{2mm}
	\sidesubfloat[]{\includegraphics[height=3.5cm]{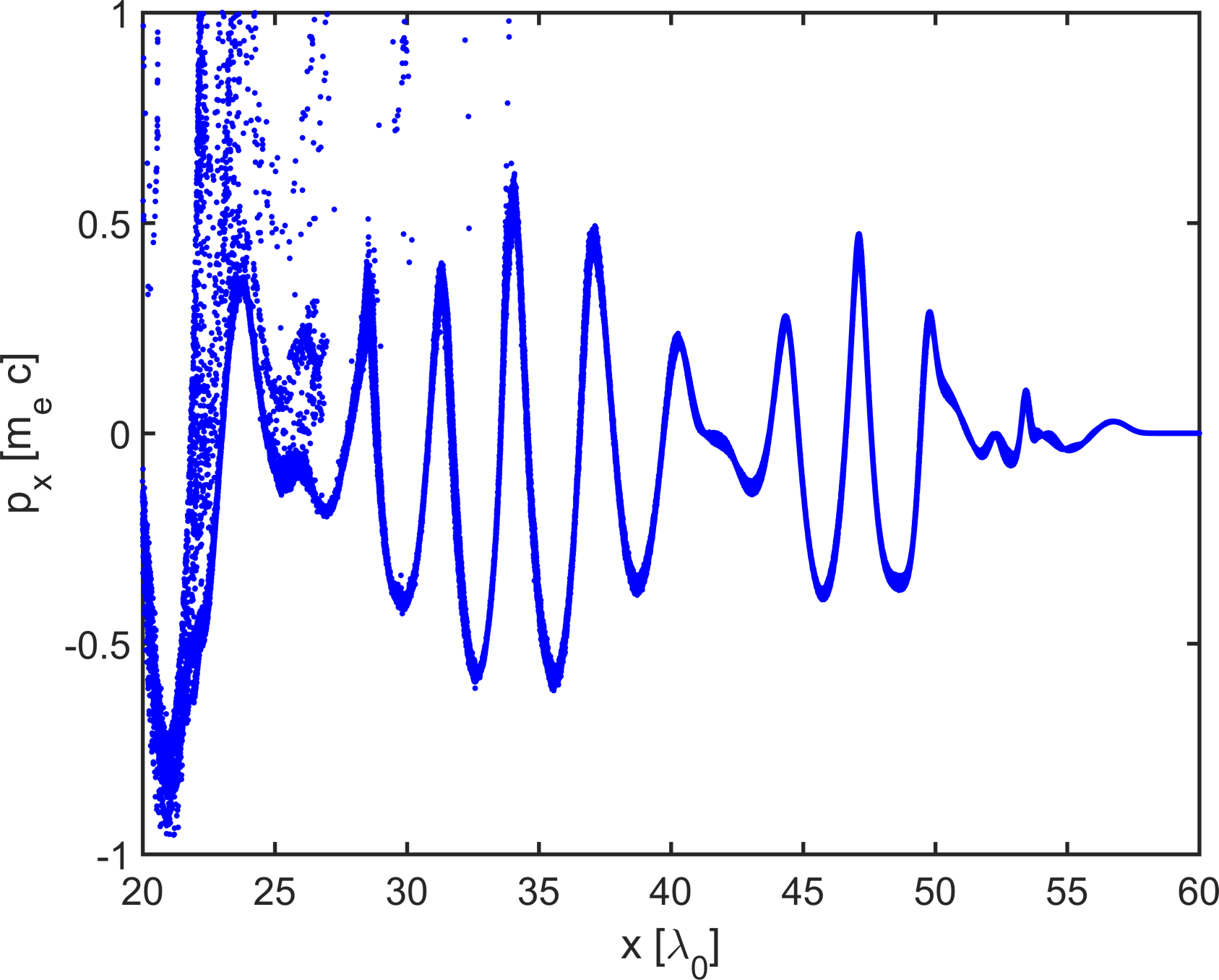}} \hspace*{5mm}
	\caption[1dlinear] {\label{fig:density} (a) Snapshot of the electron density $ n_e $, (b) its line-out along the axis $ y = 0 $ and (c) the $ (x, p_x) $ phase space at the time instant $ t = 60 \ T_0 $.}
\end{figure}

\floatsetup[figure]{style=plain, subcapbesideposition=top}
\begin{figure}[h!]
	\centering
	\vspace{3mm}
	\sidesubfloat[]{\includegraphics[height=3.5cm]{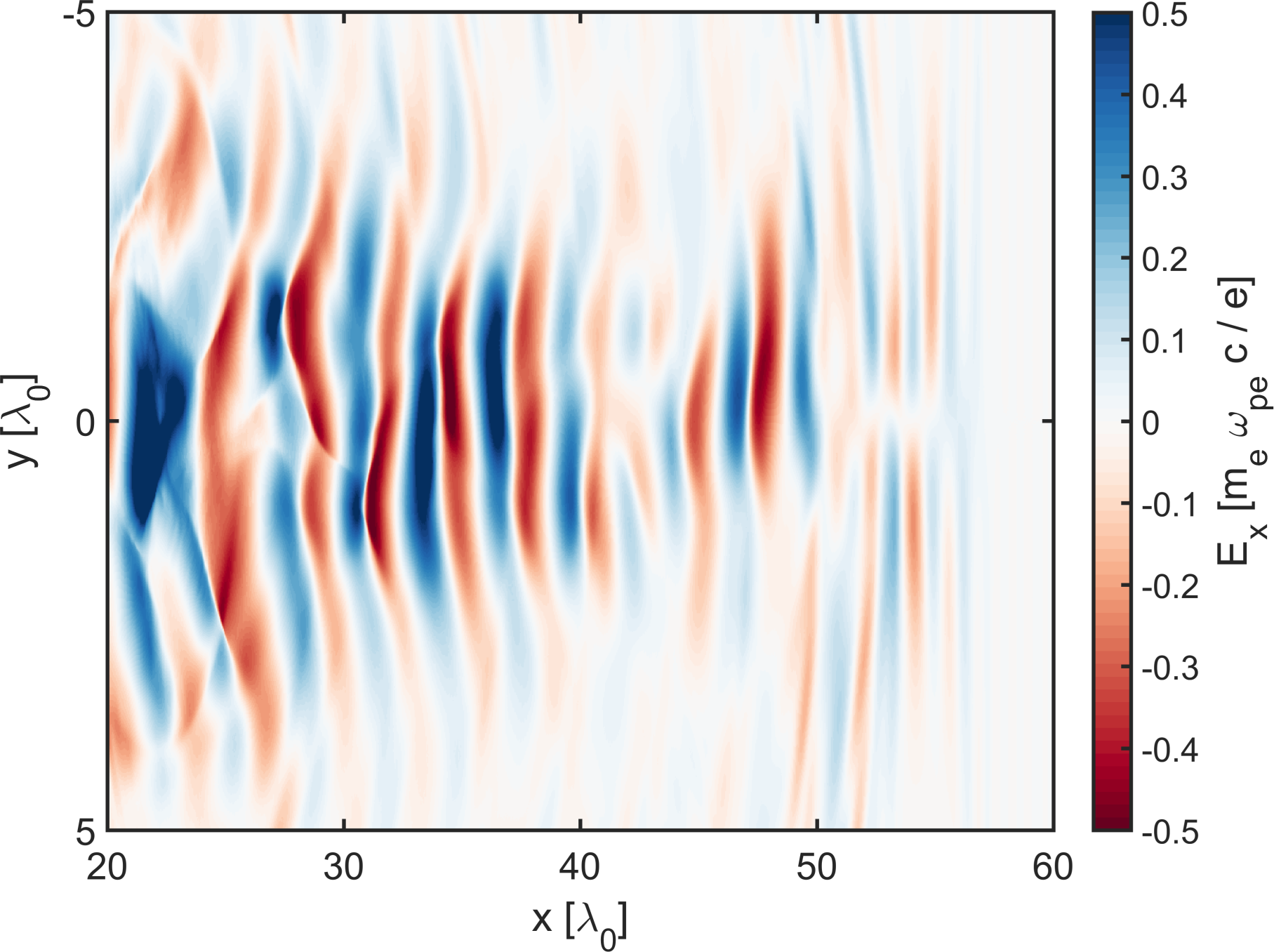}} \hspace{4mm}
	\sidesubfloat[]{\includegraphics[height=3.5cm]{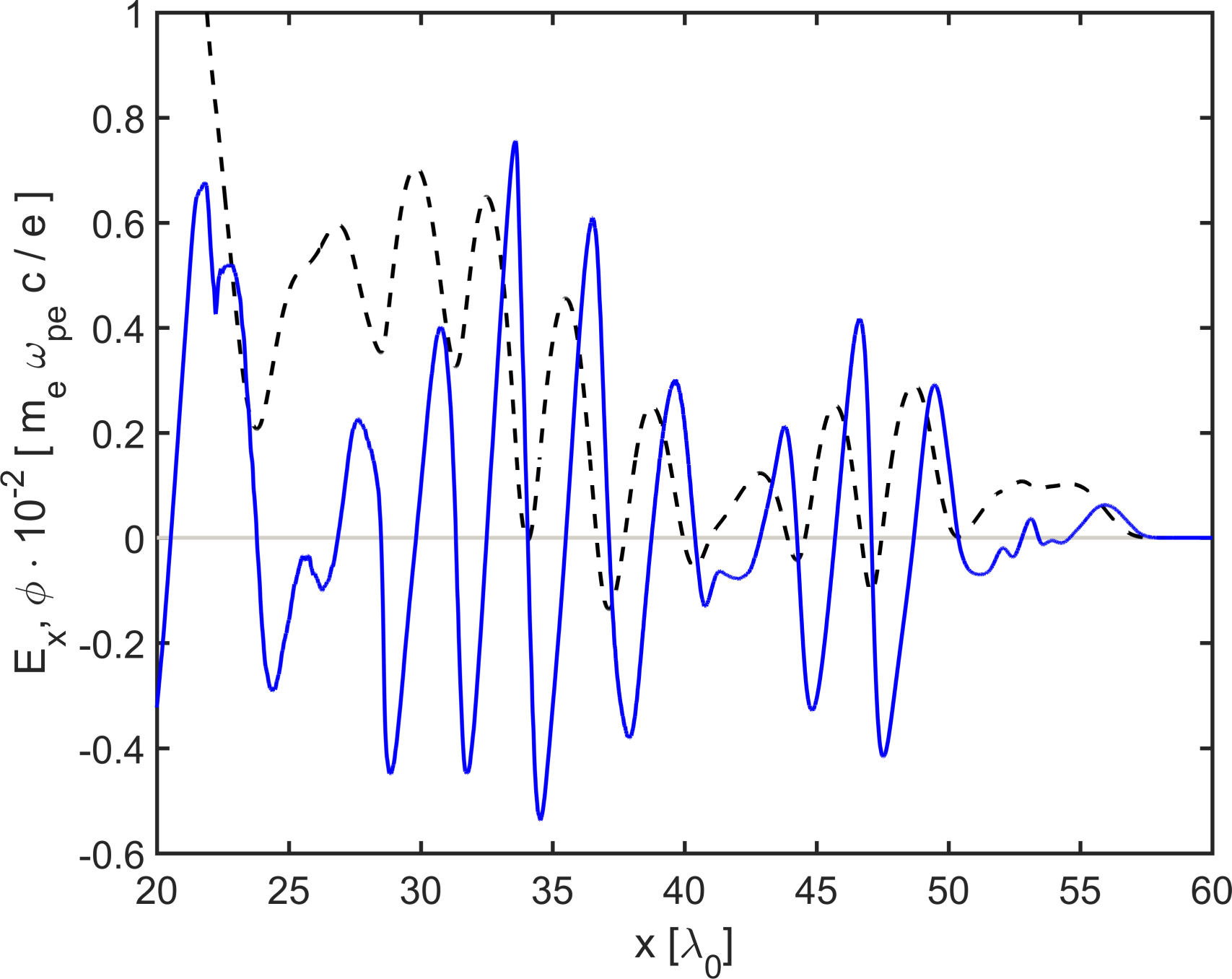}}
	\caption[1dlinear] {\label{fig:longitudinal_field} (a) Snapshot of the longitudinal electric field $ E_x $ and (b) its line-out along the axis y = 0 at the time instant $ t = 60 \ T_0 $. The dashed line in (b) represents the electric potential $ \phi $.}
\end{figure}

Due to the dispersion, the lower frequencies propagate through plasma slower than the higher frequencies which in the case of ultrashort laser pulse propagation directly results in continuous stretching of the pulse, with higher frequencies at the front and lower frequencies at the back (Fig.~\ref{fig:laser_field} (d), (e)). As can be seen in Fig.~\ref{fig:laser_field} (g) and (h), this effect, which starts to act immediately after the laser enters the density plateau at $ x = 20 \ \lambda_0 $, leads at later times to splitting of the beam into two parts propagating at different velocities. Subsequently, the gap between them further increases and the slower part may eventually remain trapped in the plasma forming a solitonlike electromagnetic modes \cite{Bulanov1992}.

Looking at the spectrum of the laser pulse (Fig.~\ref{fig:laser_field} (c), (f), (i)), one can see that the lower frequencies emerge at the beginning of the interaction and that the higher frequencies are continuously suppressed. This indicates that the central laser frequency shifts towards the lower frequencies and the group velocity of the laser pulse decreases. Consequently, the excitation of wake waves may be strongly affected since their phase velocity is in uniform plasmas proportional to the group velocity of the driving laser pulse.

To investigate the laser pulse spectrum in a greater detail, we plot the average on-axis frequency $ \left\langle \omega \right\rangle $ of the laser pulse evolving in time (Fig.~\ref{fig:density_evolution} (a)). It can be clearly seen that the down-shift of average frequency is very fast (it reaches $ 0.6 \ \omega_0 $ after only $ 20 \ \lambda_0 $ of propagation through high-density plasma region). The space-time evolution of the on-axis electron density $ n_e $ is shown in Fig.~\ref{fig:density_evolution} (b). Besides the expected deceleration of the wake waves, one can observe periodic wake wave density modulations (at $ x \approx 40 \ \lambda_0 $, $ x \approx 54 \ \lambda_0 $, $ x \approx 63 \ \lambda_0 $ and $ x \approx 68 \ \lambda_0 $). The individual modulations are localized in space for a relatively long time (see Fig.~\ref{fig:density_evolution} (c)) and their period shortens with the phase velocity of the wake waves which is in accordance with Eq.~(\ref{disp-eq}). 

The estimated time evolution of the normalized on-axis phase velocity $ \beta_w $ of the density amplitude of one selected wake wave and the corresponding $ \gamma_w $ is shown in Fig.~\ref{fig:beta_and_gamma} (a) and (b), respectively. Alongside the observed rapid decrease of the phase velocity to sub-relativistic values ($ \beta_w \approx 0.5 $ at $ x \approx 70 \ \lambda_0 $), one may also spot several local minima located at the wake wave modulations. The first modulation occurring at $ x \approx 40 \ \lambda_0 $ is likely to be the most visible, hence we will focus on its effect in further analysis.

In Fig.~\ref{fig:density} (a) and (b) one can see snapshot of the electron density $ n_e $ and its line-out along the axis $ y = 0 $, respectively, at the time instant $ t = 60 \ T_0 $. Although the structure of the wake waves is not regular mainly due to the laser polarization, we find that their on-axis amplitude drops at $ x \approx 40 \ \lambda_0 $. The same effect may be observed in the $ (x, p_x) $ phase space (Fig.~\ref{fig:density} (c)). In fact, the snapshot of the wakefield $ E_x $ (Fig.~\ref{fig:longitudinal_field} (a)) as well as its line-out along the axis $ y = 0 $ (Fig.~\ref{fig:longitudinal_field} (b)) at the time instant $ t = 60 \ T_0 $ qualitatively correspond to the space-time wakefield distribution derived in Sec.~\ref{sec:dispersion}. Therefore, modulations observed in the phase space and the plasma density distribution of wake waves is a direct consequence of the wakefield modulations caused by the effects of laser pulse dispersion and carrier envelope phase. However, the potential impact of these modulations on fast electrons remains a subject of further investigation.

\section{CONCLUSION}

In this paper, we study the propagation of few-cycle laser pulses through high-density plasmas. First, we derive analytical formulas describing the properties of the wakefield excited by few-cycle laser pulse and then we demonstrate the results of 2D PIC simulations on single-cycle laser beam propagation through near-critical density plasma. We show that the complex laser pulse evolution is dominated by the effects of dispersion and carrier envelope phase which result in the space and time modulations of the wakefield with period determined by the laser group velocity. We note that these modulations may affect trajectories of fast electrons located in the wake. Therefore, further investigation may lead to enhancement of the quality of electron sources under the conditions required for the LWFA using lasers operating at high repetition rates.

\acknowledgments 

\noindent This work was supported by the project High Field Initiative (CZ.02.1.01/0.0/0.0/15\_003/0000449) from the European Regional Development Fund.

\noindent This work was supported by The Ministry of Education, Youth and Sports from the Large Infrastructures for Research, Experimental Development and Innovations project \textquotedblleft IT4Innovations National Supercomputing Center – LM2015070\textquotedblright.

\noindent The development of the EPOCH code was funded in part by the UK EPSRC grants EP/G054950/1, EP/G056803/1, EP/G055165/1 and EP/M022463/1.

\bibliography{report} 
\bibliographystyle{spiebib} 

\end{document}